\documentstyle[11pt]{article}

\columnwidth\textwidth

\title{Gauge independence of the S-matrix in the causal approach}
\author{Andreas Aste, G\"unter Scharf \\Institut f\"ur Theoretische Physik der
Universit\"at Z\"urich, \\ Winterthurerstr. 190, CH-8057 Z\"urich, Switzerland
\and Michael D\"utsch\thanks{Work supported by Alexander von Humboldt
Foundation}\\II.Institut f\"ur Theoretische Physik, Universit\"at Hamburg\\
Luruper Chaussee 149, D-22761 Hamburg, Germany}

\date{September 30, 1997} 
\begin{document} \maketitle \begin{abstract}  The
gauge dependence of the time-ordered products for Yang-Mills theories is
analysed in perturbation theory by means of the causal method of Epstein and
Glaser together with perturbative gauge invariance.  This approach allows a
simple inductive proof of the gauge independence of the physical S-matrix.
\end{abstract} \newpage 

\def\d{\partial}\def\=d{\,{\buildrel\rm def\over =}\,}
\def\dh{\mathop{\vphantom{\odot}\hbox{$\partial$}}} \def\dl{\dh^\leftrightarrow}
\def\sqr#1#2{{\vcenter{\vbox{\hrule height.#2pt\hbox{\vrule width.#2pt
height#1pt \kern#1pt \vrule width.#2pt}\hrule height.#2pt}}}}
\def\w{\mathchoice\sqr45\sqr45\sqr{2.1}3\sqr{1.5}3\,} \def\eps{\varepsilon}
\def\oe{\overline{\rm e}} \def\onu{\overline{\nu}}
\def\ds{\hbox{\rlap/$\partial$}} \def\psq{{\overline{\psi}}}
\def\la{{(\lambda)}} \def\lap{\bigtriangleup\,}
\def\ind{(2\pi)^{-3/2}\int{d^3p\over\sqrt{2\omega}}\,} \def\Ran{{\rm Ran}\,}
\def\Ker{{\rm Ker}\,} \def\sgn{{\rm sgn}\,}

\section{Introduction}

The properties of (quantum) gauge invariance and gauge-parameter independence,
which are inherent in all kinds of gauge theories, have always been of great
interest. In the calculation of physical observables, i.e. $S$-matrix elements,
the question of gauge parameter independence arises automatically.  In the usual
Lagrangean approach to quantum field theory, the gauge invariance of the
classical Lagrangean has to be broken in order to quantize the theory.
Therefore, gauge fixing terms which depend on free gauge parameters are added to
the Lagrangean. The theory then still has BRS invariance [1]. The gauge
parameters drop out in $S$-matrix elements between physical states.  But Green
functions {\em{are}} gauge dependent in general.  On the other hand, it can be
shown that Green functions of the special class of gauge invariant operators are
independent of the method of gauge fixing and then gauge-parameter independent
[2].

In fact, the crucial property of gauge theories which allowed to show the gauge
independence of physical $S$-matrix elements by path integral methods is BRS
invariance, holding for arbitrary gauge parameters. BRS invariance implies
generalized Ward-Takahashi identities first proved by Slavnov and Taylor [3,4].
One considers then the generating functional $W_\lambda (J)$ of the theory,
where $\lambda$ is a gauge parameter and $J$ the external source coupled to a
physical field (e.g. a gauge or quark field). Changing the gauge parameter
$\lambda$ by an infinitesimal amount $d\lambda$ and using the Slavnov-Taylor
identities, the desired result can easily be derived [5].

The property of gauge parameter independence has recently gained renewed
interest also in practical problems. For example, the introduction of running
couplings can only be achieved by a resummation of certain subsets of Feynman
diagrams [6,7,8], and it is then necessary to define a general procedure for
maintaining the gauge independence of the theory.  Of course, the significance
of such resummed objects is always questionable. Furthermore, the problem has
also been discussed in the framework of the background-field model for the
electroweak Standard Model [9].

It is the aim of this paper to describe the situation from a totally different
point of view for the example of pure Yang-Mills theory, without making
reference to path integral methods. Some years ago, some of us [10, 11] began to
advocate the causal approach to perturbative quantum field theory, which goes
back to a classical paper by H. Epstein and V. Glaser [12]. No ultraviolet
divergences and only well-defined objects (no interacting fields) appear in this
approach. Meanwhile, the method has been applied successfully to full Yang-Mills
and massive theories as the electroweak Standard Model [13].

In the causal approach, the $S$-matrix is viewed as an operator-valued
distribution of the following form: $$ S(g)={\bf 1}+\sum
\limits_{n=1}^{\infty}{1\over n !} \int dx_1\ldots dx_n T_n(x_1,...x_n) g(x_1)
\cdot\ldots  g(x_n), \eqno(1.1)$$ where $g\in{\cal{S}}$, the Schwartz space of
functions of rapid decrease. The $T_n$ are well defined time-ordered products of
the first order interaction $T_1$, which specifies the theory.  For example, for
QCD without matter fields one has $$ T_1(x)=igf_{abc} \{\frac{1}{2}: A_{\mu
a}(x) A_{\nu b}(x) F^{\nu \mu}_c (x): -:A_{\mu a}(x) u_b (x) \partial^{\mu}
{\tilde{u}}_c(x): \} \quad , \eqno(1.2)$$ where $F_a^{\nu \mu} = \partial^\nu
A^\nu_a - \partial^\mu A^\nu_a$ is the {\it free} field strength tensor and
$u_a, \tilde{u}_a$ are the (fermionic) ghost fields.  The asymptotic free fields
satisfy the well-known commutation relations $$
[A_\mu^{(\pm)}(x),A_\nu^{(\mp)}(y)]=ig^{\mu \nu} D^{(\mp)}(x-y) + i
\frac{1-\lambda}{\lambda}(\partial_\mu \partial_\nu E)^{(\mp)}(x-y) \eqno(1.3)$$
in the so-called $\lambda$-gauges, and
$$\{u^{(\pm)}(x),\tilde{u}^{(\mp)}(y)\}=-iD^{(\mp)}(x-y), \eqno(1.4)$$ where D
and E will be defined below, and all other \{anti-\}commutators vanish. (For the
generalization to the massive case see [14].) The introduction of ghost fields
is necessary already at first order to preserve perturbative quantum gauge
invariance, which we are going to explain now.  It can be written in our case by
the help of an appropriately defined gauge charge $Q$: $$Q:=\lambda \int d^3x
\partial_\mu A^\mu (x)  {\dl}_0 u(x).\eqno(1.5)$$ This leads to the following
gauge variations for the fields: $$[Q,A_\mu]=i\partial_\mu u \quad , \quad
[Q,F_{\mu \nu}]=0 \quad , \quad \{Q,u\}=0 \quad , \quad \{Q,\tilde{u}\}=
-i\lambda\partial_\mu A^\mu \quad .  \eqno(1.6)$$ Obviously, these variations
have a simpler structure than those in the BRS case. Perturbative quantum gauge
invariance is then expressed by the following condition: $$ [Q,T_n(x_1,...x_n)]
= i \sum_{l=1}^{n} \partial_\mu^{x_l} T_{n/l}^{\mu} (x_1,...x_n) = ({\rm sum \,
of \, divergences}) \quad , \eqno(1.7)$$ where $T^\mu_{n/l}$ is a mathematically
rigorous version of the time-ordered product $$T^\mu_{n/l}(x_1,...,x_n) \, "="
\, T(T_1(x_1)...T^\mu_{1/1} (x_l)...T_1(x_n)) \quad ,\eqno(1.8)$$ $$[Q,T_1(x)]=:
i \partial_\nu T_{1/1}^\nu(x) \quad ,\eqno(1.9)$$ constructed by means of the
method of Epstein and Glaser.  Note that the usual 4-gluon term is missing in
$T_1$. This term is generated by quantum gauge invariance at second order of
perturbation theory.

The paper is organized as follows. In the next section we introduce the
asymptotic gauge fields in the covariant $\lambda$-gauges and discuss their
relation for different $\lambda$. Then in the third section we construct a
concrete representation in momentum space which is useful for certain
computations in the next section. There we discuss the physical subspace and
prove gauge independence of the time-ordered products on the physical subspace.
This result is a consequence of gauge invariance for any $\lambda$. The latter
property is investigated in section 5. Two appendices contain technical details.

\section{Asymptotic fields in covariant $\lambda$-gauges}

 We shall use asymptotic gauge fields satisfying the modified wave equation $$\w
A_\mu^{(\lambda)}=(1-\lambda)\d_\mu\d^\nu A_\nu^{(\lambda)}.  \eqno(2.1)$$ Here
$\lambda$ is a real gauge parameter, $\lambda=1$ corresponds to the Feynman
gauge. We have omitted color indices etc. which are unimportant in this section,
only the Lorentz structure matters here. The upper index $(\lambda)$ indicates
that the field corresponds to the gauge parameter $\lambda$, and we are going to
consider the fields with different $\lambda$ simultaneously.

First we want to solve the Cauchy problem for (2.1) with Cauchy data specified
at time $t=0$ in the whole ${\bf R^3}$. For this reason we isolate the highest
time derivatives in (2.1) $$\lambda\d_0^2A_0^\la=\lap
A_0^\la+(1-\lambda)\d_0\d^jA_j^\la \eqno(2.2)$$ $$\d_0^2A_j^\la=\lap
A_j^\la+(1-\lambda)\d_j(\d^0 A_0^\la+\d^l A_l^\la).  \eqno(2.3)$$ Consequently,
in agreement with the ordinary wave equation, the Cauchy data are given by
$A_\mu^\la (0,\vec x)$ and $(\d_0 A_\mu^\la)(0,\vec x)$. Taking the divergence
$\d^\mu$ of (2.1) we get for $\lambda \neq 0$ $$\w\d^\mu
A_\mu^\la=0,\eqno(2.4)$$ so that $A_\mu^\la$ satisfies the iterated wave
equation $$\w^2 A_\mu^\la=0.\eqno(2.5)$$ The Cauchy problem for this equation is
considered in the Appendix. The solution can be written in terms of the Lorentz
invariant distributions $D(x)$ and $E(x)$:
$$A_\mu^{(\lambda)}(x)=\int\limits_{y_0=0} d^3y\, D(x-y){\dl}_0^yA_\mu^\la
(y)+\int\limits_{y_0=0} d^3y\, E(x-y){\dl}_0^y\w A_\mu^\la(y).\eqno(2.6)$$ Here,
all second and third order time derivatives under the last integral must be
expressed by spatial derivatives of the Cauchy data by means of (2.2) and (2.3).

It is very important to notice that the decomposition (2.6) is Lorentz
covariant. Indeed, instead of selecting the plan $y_0=0$ we may consider a
smooth space-like surface $\sigma$ with a surface measure $d\sigma_\nu(y)$.
Then, with help of Gauss' theorem, the integrals in (2.6) can be written in
invariant form
$$\int\limits_{y_0=0}d^3y\ldots{\dl}_0\rightarrow\int\limits_\sigma
d\sigma_\nu(y)\ldots{\dl}^\nu,$$ showing that each term on the r.h.s. of (2.6)
is a Lorentz four-vector.

Let us denote the first term in (2.6) which satisfies the ordinary wave equation
by $A_\mu^w(x)$ ($w$ for wave). The second term denoted by $B_\mu$ is equal to
$$B_\mu(x)=(1-\lambda)\int d^3y\,E(x-y){\dl}_0\d_\mu\d A^\la(y),\eqno(2.7)$$
where (2.1) has been inserted. For $\mu=j=1,2,3$ the derivative can be taken out
by partial integration, so that $$B_j(x)=\d_j\chi(x)\eqno(2.8)$$ is a spatial
gradient with $$\chi(x)=(1-\lambda)\int\limits_{y_0=0}d^3y\,E(x-y){\dl}_0\d
A^\la(y).  \eqno(2.9)$$ But this is impossible for the zeroth component
$$B_0(x)=\d_0\chi(x)+B(x).\eqno(2.10)$$ The difference $B(x)$ can be transformed
as follows $$B(x)=(1-\lambda)\int d^3y\,[E(x-y){\dl}_0\d_0^y\d A^\la (y)-\d_0^x
E(x-y) {\dl}_0\d A^\la (y)]$$ $$=(1-\lambda)\int d^3y\,\d_0^y[E(x-y){\dl}_0\d
A^\la (y)]$$ $$=(1-\lambda)\int d^3y[\d_0^yE\d_0\d A^\la+E\d_0^2\d
A^\la-\d_{0y}^2E\d A^\la- \d_0^yE\d_0\d A^\la]$$ $$=-(1-\lambda)\int d^3y\,\w
E(x-y)\d A^\la (y)=-(1-\lambda)\int d^3y\, D(x-y)\d^\nu A_\nu^\la
(y).\eqno(2.11)$$

This shows that the field $B(x)$ also fulfills the wave equation $\w B=0$.
Therefore, it is tempting to combine it with $A_0^w(x)$. The resulting
four-component field $$A_\mu^L=(A_0^w+B,A_1^w,A_2^w,A_3^w)\eqno(2.12)$$
satisfies the wave equation and we have the simple decomposition $$A_\mu^\la
=A_\mu^L+\d_\mu\chi.\eqno(2.13)$$ However, this decomposition has the serious
defect of not being covariant (see (2.12)). Therefore, we must make a sharp
distinction between the field $A_\mu^L$ and the {\it covariant} field $A_\mu^F$
in the Feynman gauge $\lambda=1$, although both fields satisfy the wave equation
and the same commutation relation, as we shall see.

Next we want to quantize the $A^\la$-field. It follows from (2.6) that the
commutation relations for arbitrary times must involve the distributions $D$ and
$E$. Then, Poincar\'e covariance and the singular order $\omega=-2$ of the
resulting distribution suggest the following form
$$[A_\mu^\la(x),\,A_\nu^\la(y)]=ig_{\mu\nu}D(x-y)+i\alpha\d_\mu\d_\nu
E(x-y),\eqno(2.14)$$ where a common factor ($h/2\pi$) has been set =1. When
operating with $\w g^{\kappa \mu}-(1-\lambda)\d^\kappa\d^\mu$ on the variable
$x$, we must get zero. This determines the parameter $\alpha$. Using $\w
E(x)=D(x)$, we find
$$[A_\mu^\la(x),\,A_\nu^\la(y)]=ig_{\mu\nu}D(x-y)+i{1-\lambda
\over\lambda}\,\d_\mu\d_\nu E(x-y).\eqno(2.15)$$ The corresponding commutation
relations for the positive and negative frequency parts read (see (1.3))
$$[A_\mu^{\la(-)}(x),\,A_\nu^{\la(+)}(y)]=ig_{\mu\nu}D^{(+)}(x-y)+
i{1-\lambda\over\lambda}\,(\d_\mu\d_\nu E)^{(+)}(x-y).\eqno(2.16)$$ Note that
the positive frequency part of the derivative $(\d_\nu E)$ is well defined, in
contrast to $E^{(+)}$.

>From the initial values of the $D$- and $E$-distributions $$D(0,\vec x)=0,\quad
(\d_0D)(0,\vec x)=\delta^3(\vec x)\eqno(2.17)$$ $$(\d_0^nE)(0,\vec x)=0,\quad
n=0,1,2,\quad (\d_0^3E)(0,\vec x)=\delta^3(\vec x)\eqno(2.18)$$ we obtain the
equal-time commutation relations
$$[\d_0A_\mu^\la(x),A_\nu^\la(y)]_0=ig_{\mu\nu}\Bigl(1+g_{\mu 0}
{1-\lambda\over\lambda}\Bigl)\delta(\vec x-\vec y)\eqno(2.19)$$
$$[\d_0A_0^\la(x),\d_0A_j^\la(y)]_0=i {\lambda-1\over\lambda}\d_j\delta(\vec
x-\vec y),\eqno(2.20)$$ where the subscript 0 means $x_0=y_0$. All other
commutators are zero.  It follows from (2.19) (2.20) that 3-dimensional smearing
with a space dependent test function $f(\vec x)$ is sufficient to get a
well-defined operator in Fock space.

>From the fundamental commutation relations (2.15) the commutators of all other
fields can be calculated because they are all expressed by $A_\mu^\la$. We find
$$[\chi(x),\chi(y)]=0\eqno(2.21)$$
$$[A_j^w(x),A_k^w(y)]=ig_{jk}D(x-y)\eqno(2.22)$$
$$[A_0^w(x),A_0^w(y)]={i\over\lambda}D(x-y)\eqno(2.23)$$
$$[A_\mu^L(x),A_\nu^L(y)]=ig_{\mu\nu}D(x-y).\eqno(2.24)$$ Now, $A_j^w(x),
j=1,2,3$ are the spatial components of a covariant vector field satisfying the
wave equation. The commutation relations (2.22) are the same as for the Feynman
field $A_j^F(x)$. Nevertheless, we cannot identify the two as we shall see in
the next section by constructing a concrete representation of the field
operators.

\section{Concrete representation in momentum space}

Most authors who consider the $\lambda$-gauges leave the construction of a
concrete representation to the reader. We try to be more polite to our readers.

Since three-dimensional smearing is enough to render $A_\mu^\la(x)$ well
defined, we will construct all fields as three-dimensional Fourier integrals,
leaving aside manifest Lorentz covariance. Our strategy will be to start with a
representation of the time-zero fields which satisfies the equal-time
commutation relations (2.19) (2.20) and then calculating the time evolution by
the formulae of the last section. We follow the somewhat unusual, but
mathematically more satisfactory procedure of assuming a Fock space with {\it
positiv definite metric} and changing the form of the zeroth component $A_0^\la$
instead [10]. This is very natural in the $\lambda$-gauge because the zeroth
component plays a special role here, anyway.

We use the usual emission and absorption operators for all four components
satisfying $$[a_\nu^\la(\vec p),a_\mu^{\la +}(\vec
q)]=\delta_{\nu\mu}\delta(\vec p -\vec q).\eqno(3.1)$$ The adjoint is defined
with respect to the positive definite scalar product so that these operators can
be represented in the usual way in a Fock space $\cal F^\la$, if smeared out
with test functions $f(\vec p)\in L^2({\bf R^3})$. In addition we will use the
operators for the longitudinal mode $$a_\parallel^\la (\vec
p)={p^j\over\omega}a_j^\la(\vec p)=-{p_j\over\omega} a_j^\la(\vec
p),\eqno(3.2)$$ where always $\omega=|\vec p|=p^0$. Introducing the linear
combinations $$b_1^\la(\vec p)={1\over\sqrt{2}}(a_\parallel^\la(\vec p)+a_0^\la
(\vec p))$$ $$b_2^\la(\vec p)={1\over\sqrt{2}}(a_\parallel^\la(\vec p)-
a_0^\la(\vec p)),\eqno(3.3)$$ we have the following commutators $$[b_1^\la(\vec
p),b_2^{\la +}(\vec q)]=0$$ $$[a_\nu^\la(\vec p),b_2^{\la +}(\vec
q)]=-{1\over\sqrt{2}}{p_\nu \over\omega}\delta(\vec p-\vec q)$$ $$[a_\nu^{\la
+}(\vec p),b_1^\la(\vec q)]=-{1\over\sqrt{2}}{p^\nu\over \omega}\delta(\vec
p-\vec q).\eqno(3.4)$$ In addition to the adjoint we have to introduce a second
conjugation $K$ which appears in all Lorentz covariant expressions and defines
the so-called Krein structure [17]. It is defined by $$a_0^\la(\vec
p)^K=-a_0^\la(\vec p)^+,\quad a_j^\la(\vec p)^K= a_j^\la(\vec p)^+, \quad
j=1,2,3.\eqno(3.5)$$ Note that $a_\mu^\la, a_\mu^{\la }+$ are not treated as
four-vectors, therefore, we write the indices always downstairs.

The gauge field $A_\mu^\la(x)$ must be self-conjugated $A_\mu^{\la K}=
A_\mu^\la$.  Then, a little experimentation shows that the time-zero fields must
be of the following form: $$A_\mu^\la (0,\vec x)=\ind\biggl\{a_\mu^\la(\vec
p)e^{i\vec p\vec x}+ a_\mu^{\la K}(\vec p)e^{-i\vec p\vec x}$$
$$-{1-\lambda\over 2\sqrt{2}\lambda}\biggl[{p_\mu\over\omega}b_1^\la(\vec p)
e^{i\vec p\vec x}+{p_\mu\over\omega}b_2^{\la +}(\vec p)e^{-i\vec p\vec x}$$
$$-2g_{\mu 0}b_1^\la(\vec p)e^{i\vec p\vec x}- 2g_{\mu 0}b_2^{\la +}(\vec
p)e^{-i\vec p\vec x}\biggl]\biggl\},\eqno(3.6)$$ $$(\d_0A_\mu^\la)(0,\vec
x)=-i\ind\biggl\{\omega a_\mu^\la(\vec p)e^ {i\vec p\vec x}-\omega a_\mu^{\la
K}(\vec p)e^{-i\vec p\vec x}$$ $$-{1-\lambda\over 2\sqrt{2}\lambda}\biggl[-p_\mu
b_1^\la(\vec p) e^{i\vec p\vec x}+p_\mu b_2^{\la +}(\vec p)e^{-i\vec p\vec x}$$
$$-2g_{\mu 0} \omega b_1^\la(\vec p)e^{i\vec p\vec x}+ 2g_{\mu 0} \omega
b_2^{\la +}(\vec p)e^{-i\vec p\vec x}\biggl]\biggl\}. \eqno(3.7)$$ It is
straight-forward to verify the commutation relations (2.15) (2.19) (2.20).

The fields for arbitrary times can now be found from (2.6). For this purpose we
need the following three-dimensional Fourier transforms (for $y_0=0$) $$\int
d^3y\,D(x-y)e^{i\vec p\vec y}=-{i\over 2\omega}\Bigl( e^{i \omega
x^0}-e^{-i\omega x^0}\Bigl)e^{i\vec p\vec x}\eqno(3.8)$$ $$\int
d^3y\,E(x-y)e^{i\vec p\vec y}={-1\over 4\omega^2}\Bigl[ e^{i \omega
x^0}\Bigl(x^0+{i\over\omega}\Bigl)+e^{-i\omega x^0}\Bigl(
x^0-{i\over\omega}\Bigl)\Bigl]e^{i\vec p\vec x}.\eqno(3.9)$$

We first compute $$(\d^\mu A_\mu^\la)(0,\vec
x)={-i\over\lambda}(2\pi)^{-3/2}\int d^3p\,\sqrt{\omega}\Bigl(b_1^\la(\vec
p)e^{i\vec p\vec x}-b_2^{\la +} (\vec p)e^{-i\vec p\vec x}\Bigl),\eqno(3.10)$$
$$(\d_0\d^\mu A_\mu^\la)(0,\vec x)={-\sqrt{2}\over\lambda}\ind\omega^2
\Bigl(b_1^\la(\vec p)e^{i\vec p\vec x}+b_2^{\la +}(\vec p)e^{-i\vec p\vec x}
\Bigl).\eqno(3.11)$$ Then we obtain from the first term in (2.6)
$$A_j^w(x)=\ind\Bigl(a_j^\la(\vec p)e^{-ipx}+a_j^{\la +}(\vec p)e^{ipx}\Bigl) +
\d_j f(x) \quad, \eqno(3.12)$$ where $$ f(x)=-i\frac{1-\lambda}{4 \lambda (2
\pi)^{3/2}} \int \frac{d^3 p}{\omega^ {3/2}} \Bigl[ b_1^\la(-\vec p) e^{ipx} -
b_2^{\la +} (-\vec p) e^{-ipx} \Bigr].$$ The zeroth component behaves
differently $$A_0^w(x)=\ind\Bigl[a_0^\la(\vec p)e^{-ipx}-a_0^{\la +}(\vec
p)e^{ipx}$$ $$+{1-\lambda\over\sqrt{2}\lambda}\Bigl(b_1^\la(\vec p)e^{-ipx}+
b_2^{\la +}(\vec p)e^{ipx}\Bigl)\Bigl]$$
$$-{1-\lambda\over{2\sqrt{2}}\lambda}\ind\Bigl(b_1^\la(-\vec p)e^{ipx} +
b_2^{\la +}(-\vec p)e^{-ipx}\Bigl) .\eqno(3.13)$$ Next we calculate $\chi(x)$
from (2.9) $$\chi(x)={1-\lambda\over\sqrt{2}\lambda}\ind\Bigl[b_1^\la(\vec p)
e^{-ipx}\Bigl(x^0-{i\over 2\omega}\Bigl)+b_2^{\la +}(\vec p)e^{ipx}\Bigl(
x^0+{i\over 2\omega}\Bigl)\Bigl]-f(x)\eqno(3.14)$$ and $B(x)$ from (2.11)
$$B(x)={\lambda-1\over\sqrt{2}\lambda}\ind\Bigl(b_1^\la(\vec p)e^{-ipx}+
b_2^{\la +}(\vec p)e^{ipx}\Bigl)$$ $$
-{\lambda-1\over\sqrt{2}\lambda}\ind\Bigl(b_1^\la(-\vec p)e^{ipx} + b_2^{\la
+}(-\vec p)e^{-ipx}\Bigl) .\eqno(3.15)$$ This cancels against the second line in
(3.13) so that $$A_0^\la(x)=\ind\Bigl[a_0^\la(\vec p)e^{-ipx}-a_0^{\la +}(\vec
p) e^{ipx}\Bigl]+\d_0(\chi+f).\eqno(3.16)$$ The first integral in (3.16) and
(3.12) formally agrees with the Feynman field $A_\mu^F$, but the latter is
defined by means of different annihilation and creation operators
$a_\mu^{(1)}(\vec p)$, $a_\mu^{(1)+}(\vec p)$
$$A_\mu^F(x)=\ind\Bigl[a_\mu^{(1)}(\vec p)e^{-ipx}+a_\mu^{(1)K}(\vec p)
e^{ipx}\Bigl].\eqno(3.17)$$ In $A_\mu^\la$ the terms with wrong frequencies
$\sim b_1^\la(-\vec p)$ etc. cancel out. Then the resulting decomposition
$A_\mu^ \la =\tilde A_\mu^L+ \d_\mu\tilde\chi$ is identical with the one
introduced by Lautrup [16].

Until now every field $A^\la$ operates in its own Fock space $\cal F^\la$. But
there must exist a $\lambda$-independent intersection of these $\cal F^\la$
where the gauge independent objects live. Indeed, in the foregoing equations the
$\lambda$-dependence is only through the unphysical scalar and longitudinal
modes $b_1, b_2$ (3.3), all equations involving only transverse modes which can
be written down contain no $\lambda$. Therefore, we can safely identify the
transverse emission and absorption  operators for different $\lambda$. Let
$\eps^\mu=(0,\vec\eps)$ and $\eta^\mu=(0,\vec\eta)$ be two transverse
polarization vectors $$\vec p\cdot\vec\eps(\vec p)=0=\vec p\cdot\vec\eta(\vec
p),\quad \vec\eps^2=1=\vec\eta^2,\quad\vec\eps\cdot\vec\eta=0.\eqno(3.18)$$ Then
we put $$\eps^\mu a_\mu^\la(\vec p)=a_\eps(\vec p),\quad \eta^\mu a_\mu^\la(\vec
p)=a_\eta(\vec p)\eqno(3.19)$$ independent of $\lambda$. Choosing one unique
vacuum $\Omega$ for all field operators $$a_\eps(\vec p)\Omega=0=a_\eta(\vec
p)\Omega=b_1^\la(\vec p)\Omega= b_2^\la(\vec p)\Omega=0$$ for all $\vec p$ (or
rather after smearing with test functions $f(\vec p)$), then the different Fock
spaces $\cal F^\la$ hang together (fig.1). Their intersection is the physical
subspace $\cal H_{\rm phys}$ which is spanned by the transverse states
$(a_\eps^+)^m(a_\eta^+)^n\Omega$.

\section{Gauge invariance and gauge independence}

Now we come to the study of the nilpotent gauge charge $Q_\lambda$ (1.5)
$$Q_\lambda=\lambda \int d^3x \partial^\mu A_\mu^\la (x)  {\dl}_0 u(x),
\eqno(4.1)$$ where the color indices are always suppressed if the meaning is
clear.  The ghost fields $u$, $\tilde u$ are quantized as follows $$\w u=0,\quad
\w {\tilde{u}}=0$$ $$\{u_a(x),\tilde u_b(y)\}=-i\delta_{ab}D(x-y).\eqno(4.2)$$
Since there is no  $\lambda$-dependence here, they can be represented in the
usual way $$u(x)=\ind \Bigl(c_2(\vec p)e^{-ipx}+c_1^+(\vec p)e^{ipx}\Bigl)
\eqno(4.3)$$ $$\tilde u(x)=\ind \Bigl(-c_1(\vec p)e^{-ipx}+c_2^+(\vec
p)e^{ipx}\Bigl) \eqno(4.4)$$ where $$\{c_i(\vec p),c_j^+(\vec
q)\}=\delta_{ij}\delta(\vec p-\vec q), \quad i,j=1,2.$$ The conjugation $K$ is
extended to the ghost sector by $$c_2(\vec p)^K=c_1(\vec p)^+,\quad c_1(\vec
p)^K=c_2(\vec p)^+ \eqno(4.5)$$ so that $u^K=u$ is $K$-selfadjoint and $\tilde
u^K=-\tilde u$. Then $Q_\lambda$ (4.1), if densely defined, becomes
$K$-symmetric $Q_\lambda \subset Q_\lambda^K$. It is not necessary for the
following to give an explicit description of the domain. According to a general
result [18], it has a $K$-selfadjoint extension $Q_\lambda^K=Q_\lambda$ which is
a closed operator and this is all we need for our purpose.

Using (3.10), (3.11) and (4.3) it is easy to calculate $Q_\lambda$ in momentum
space $$Q_\lambda=\sqrt{2}\int d^3p\,\omega(\vec p)[b_1(\vec p)c_1^+(\vec p)
+b_2^+(\vec p)c_2(\vec p)].\eqno(4.6)$$ For typographical simplicity we have not
written the $\lambda$-dependence in $b_1, b_2$. $Q_\lambda$ together with its
adjoint $$Q_\lambda^+=\sqrt{2}\int d^3p\,\omega(\vec p)[c_1(\vec p)b_1^+(\vec p)
+c_2^+(\vec p)b_2(\vec p)]\eqno(4.7)$$ are unbounded closed operators; the
unboundedness is not only due to the emission and absorption operators but also
because of $\omega(\vec p) = |\vec p|$.

Since $Q_\lambda$, $Q_\lambda^+$ are closed operators, we have the following
direct decompositions of the Fock space $${\cal F}^\la =\overline{\Ran
Q_\lambda}\oplus\Ker Q_\lambda^+= \overline{\Ran Q_\lambda^+}\oplus\Ker
Q_\lambda,\eqno(4.8)$$ where Ran is the range and Ker the kernel of the
operator. The overline denotes the closure; note that $\Ran Q_\lambda$ is not
closed because 0 is in the essential spectrum of $Q_\lambda$. Now,
$Q_\lambda^2=0$ implies $\Ran Q_\lambda\perp\Ran Q_\lambda^+$, therefore, it
follows from (4.8) that $${\cal F}^\la =\overline{\Ran
Q_\lambda}\oplus\overline{\Ran Q_\lambda^+}\oplus\Bigl(\Ker Q_\lambda\cap\Ker
Q_\lambda^+\Bigl).  \eqno(4.9)$$ The range of $Q_\lambda$ and $Q_\lambda^+$
certainly consists of unphysical states because (4.6) and (4.7) only contains
emission operators of unphysical particles (scalar and longitudinal "gluons" and
ghosts). The physical states must therefore be contained in the last subspace in
(4.9).

We claim that $$\Ker Q_\lambda\cap\Ker Q_\lambda^+=\Ker\{Q_\lambda,Q_\lambda^+\}
\eqno(4.10)$$ where the curly bracket is the anticommutator. Indeed, if a vector
$f\in {\cal F^\la}$ belongs to the l.h.s., that means $Q_\lambda f=0=Q_\lambda
^+ f$ then it is also contained in the r.h.s. Inversely, if $f$ belongs to the
r.h.s. then $$0=(f,\{Q_\lambda,Q_\lambda^+\}f)=\|Q_\lambda
f\|^2+\|Q_\lambda^+f\|^2, $$ it is also contained in the l.h.s. Calculating the
anticommutator from (4.6) (4.7) we find $$\{Q_\lambda,Q_\lambda^+\}=2\int
d^3p\,\omega^2(\vec p)\Bigl[b_1^+(\vec p)b_1(\vec p)+b_2^+(\vec p)b_2(\vec
p)+c_1^+(\vec p)c_1(\vec p)+c_2^+ (\vec p)c_2(\vec p)\Bigl].\eqno(4.11)$$ Up to
the (positive) factor $\omega^2$ this is just the particle number operator of
the unphysical particles. The physical subspace is characterized by the fact
that there are no unphysical particles, hence, $${\cal H}_{\rm
phys}=\Ker\{Q_\lambda,Q_\lambda^+\}\eqno(4.12)$$ and this is a closed subspace.
As discussed above, it is the intersection of all $\cal F^\la$.

We introduce the projection operator $P_\lambda$ on ${\cal H}_{\rm phys}$.  It
is our goal to prove the gauge independence of the physical S-matrix $P_\lambda
S^\la(g)P_\lambda$. The perturbative formulation in terms of time-ordered
products $T_n^\la$ (1.1) would be $$P_\lambda T_n^\la P_\lambda=P_1 T_n^{(1)}P_1
+{\rm div.}\eqno(4.13)$$ Here div denotes a sum of divergences which vanish
after integration with test functions $g(x_1)$ $\ldots$ $g(x_n)$ in the formal
adiabatic limit where terms with derivatives of $g$ are neglected.  In (4.13) we
 have compared the physical $n$-point functions in the $\lambda$-gauge with the
Feynman gauge $\lambda=1$.

Gauge independence (4.13) is a direct consequence of gauge invariance (1.7).
That (1.7) really holds for arbitrary $\lambda$ is discussed in the next
section. Gauge invariance implies the following important proposition (see [14],
eq.(5.28)) $$PT(X_1)PT(X_2)P=PT(X_1)T(X_2)P+{\rm div}.\eqno(4.14)$$ Here we have
omitted indices $n$ and subscripts $\lambda$ to indicate that (4.14) holds for
arbitrary $\lambda$ and arbitrary $n$-point functions. For the sake of
completeness we give a proof of (4.14) in Appendix 2.

The proof of gauge independence is by induction on $n$. The beginning $n=1$ can
be easily verified because $P_\lambda A_\mu^\la P_\lambda = P_1 A_\mu^{(1)}P_1$
and $T_1$ (1.2) does not depend explicitly on $\lambda$. Let us now assume that
$$P_\lambda T_i^\la P_\lambda=P_1 T_i^{(1)} P_1+{\rm div}\eqno(4.15)$$ holds for
all $i\le n-1$. Then we consider arbitrary products $$P_\lambda
T^\la(X_1)T^\la(X_2)P_\lambda =P_\lambda T^\la(X_1)P_\lambda
T^\la(X_2)P_\lambda+{\rm div}_1, \eqno(4.16)$$ where we have used (4.14). Due to
the induction assumption (4.15) this is equal to $$=P_1 T^{(1)}(X_1)P_1
T^{(1)}(X_2)P_1 +{\rm div}_2= P_1 T^{(1)}(X_1)T^{(1)}(X_2)P_1 +{\rm
div}_3.\eqno(4.17)$$ Here we have used (4.14) again. The causal $D$-distribution
of order $n$ in the Epstein-Glaser construction is a sum of such products
(4.16), hence, it follows that $$P_\lambda D_n^\la P_\lambda=P_1
D_n^{(1)}P_1+{\rm div}. \eqno(4.18)$$ All three terms in here have separately
causal support, therefore they can individually be split into retarded and
advanced parts. The local nomalization terms can be chosen in such a way that
$$P_\lambda R_n^\la P_\lambda=P_1R_n^{(1)}P_1+{\rm div},\eqno(4.19)$$ where $R$
denotes the retarded distributions. We must check that this way of normalization
is not in conflict with the normalization which we adopt to achieve gauge
invariance (see next section). But this is not the case for the following
reason. We decompose $$T_n=PT_nP+W_n.$$ The condition (4.19) concerns the
physical part $PT_nP$, only.  But the latter is gauge invariant for any
normalization $$QPT_nP-PT_nPQ=0$$ because $PQ=0=QP$. Therefore, the
normalization in the proof of gauge invariance involves only the unphysical part
$W_n$.  From  the gauge independence of the retarded distributions (4.19) we get
the same result for the $n$-point distributions $$P_\lambda T_n^\la
P_\lambda=P_1T_n^{(1)}P_1+{\rm div}\eqno(4.20)$$ in the usual way. This
completes the inductive proof.

\section{Gauge invariance in an arbitrary $\lambda$-gauge}

Gauge invariance (1.7) has been proven in the Feynman gauge $\lambda =1$
[11,15,19]. Here we summarize and reformulate that proof in a way which is
manifestly independent of the choice of $\lambda$.

The generator $Q_\lambda$ (4.1) of the gauge transformations depends on
$\lambda$. But the parameter $\lambda$ drops out in the gauge variations of the
free fields $A^\mu,\,F^{\mu\nu}$ and $u$ with the exception of $\tilde u$ (1.6).
However, we shall work with a ghost coupling (1.2) containing the field $\tilde
u_a$ in the form $\d_\mu\tilde u_a$ only. The gauge variation of the latter
field can be written in a $\lambda$-independent form $$\{Q,\d_\mu\tilde
u_a\}=-i\lambda\d_\mu\d^{\nu}A_{a\nu} =i\d^\nu F_{a\nu\mu}\eqno(5.1)$$ by means
of the equation of motion (2.1).  \vskip 0.5cm {\it 1. Gauge invariance at first
order } \vskip 0.5cm The coupling (1.2) is gauge invariant at first order (1.9)
with the 'Q-vertex' $$T^\nu_{1/1}(x)\=d igf_{abc}[:A_{\mu a}(x)u_b(x)F_c^{\nu
\mu}(x): -{1\over 2}:u_a(x)u_b(x)\d^\nu\tilde u_c(x):].\eqno(5.2)$$ for any
value of the gauge parameter $\lambda$.  The most general coupling which is
gauge invariant at first order, symmetrical (Lorentz covariant, SU(N)-invariant,
P-,T-,C-invariant, pseudo-unitary) and is compatible with renormalizability
contains a non-uniqueness in the ghost sector [21]
$$T_1+\beta_1\{Q,gf_{abc}:u_a\tilde u_b\tilde u_c:\} +\beta_2\d_\mu
[igf_{abc}:A^\mu_a u_b\tilde u_c:],\eqno(5.3)$$ $\beta_1,\beta_2\in{\bf R}$
arbitrary, and $T_1$ is given by (1.2). We shall prove gauge invariance in the
case $\beta_1=0=\beta_2$. For general values of $\beta_1,\beta_2 \in{\bf R}$ a
manifestly $\lambda$-independent formulation is impossible, since the fields
$\tilde u_a$ (without derivative) (1.6) and $\d_\mu A^\mu_a$ appear in the
coupling. (The 'bad' behavior of  $\d_\mu A^\mu_a$ is explained below in
subsect.5.3) But it has been proven [20] that gauge invariance for
$\beta_1=0=\beta_2$ implies gauge invariance for arbitrary
$\beta_1,\beta_2\in{\bf R}$ at least at low orders. The argumentation of that
proof is of general kind, such that it applies to any choice of $\lambda$.
\vskip 0.5cm {\it 2. Outline of the proof of gauge invariance in feynman gauge
$\lambda=1$} \vskip 0.5cm In this subsect. we summarize the proof of gauge
invariance (1.7) which was given for $\lambda=1$ in [11,15,19]. In the next
subsect. we shall see that this proof needs no modifications for arbitrary
$\lambda$.  The proof is by induction on the order $n$ of the perturbation
series.  The operator gauge invariance (corresponding to (1.7)) of $A^\prime_n,
R^\prime_n$ and $D_n= A^\prime_n-R^\prime_n$,
$$[Q,D_n(x_1,...,x_n)]=i\sum_{l=1}^n\d_\mu^{x_l}D^
\mu_{n/l}(x_1,...,x_n),\eqno(5.4)$$ has been proven in a straightforward way
[11] from the gauge invariance of the $T_m, \, m \leq n-1$.  This proof is very
instructive because it shows that our definition (1.7) of gauge invariance is
adapted to the inductive construction of the $T_n$'s.  However, the distribution
splitting $D_n=R_n-A_n$ can only be done in terms of the numerical distributions
$d_n=r_n-a_n$. Therefore, we have to express the operator gauge invariance (5.4)
by the {\it Cg-identities} for $D_n$, the C-number identities for gauge
invariance, which imply the operator gauge invariance (5.4).

However, there is a serious problem [11,19]. Terms with different field
operators may compensate, due to identities like $$[u(x_1)-u(x_2)]\d^{x_1}_\mu
\delta (x_1-x_2)+\delta (x_1-x_2)\d_\mu u(x_1) =0.\eqno(5.5)$$ Therefore the
definition of the C-number distributions in $R_n^\prime, A_n^\prime$ (and
therefore in $D_n=R_n^\prime-A_n^\prime$) has a certain ambiguity because terms
$\sim (\d\delta):A...:$ can mix up with terms $\sim \delta:\d A...: \sim
\delta:F...:.$ To get rid of these ambiguities, we choose the convention of {\it
only} applying Wick's theorem (doing nothing else) to
$$A'_n(x_1,...;x_n)=\sum_{Y,Z} \tilde T_k (Y) T_{n-k}(Z,x_n),\eqno (5.6)$$ where
the (already constructed) operator decompositions of $\tilde T_k,\,T_{n-k}$ are
inserted. In this way we obtain the so-called {\it natural} operator
decomposition of $A'_n$ $$A'_n=\sum_{\cal O} a'_{\cal O}:{\cal O}:,\eqno(5.7)$$
where the sum runs over all combinations ${\cal O}$ of free field operators. We
similarly proceed with $A^\prime_{n/l}, R^\prime_n$ and $R^\prime_{n/l}$ and
define $d_{\cal O}^{(l)}\=d r_{\cal O}^{\prime (l)}-a_{\cal O}^{\prime (l)}$.
Then, we split the numerical distributions $d_{\cal O}^{(l)}$ with respect to
their supports into retarded and advanced parts $d_{\cal O}^{(l)}= r_{\cal
O}^{(l)}-a_{\cal O}^{(l)}$. Next we define $t_{\cal O} ^{\prime (l)}\=d r_{\cal
O}^{(l)}-r_{\cal O}^{\prime (l)}$ and symmetrize it, which yields  $t_{\cal
O}^{(l)}$. The definition $$T_{n(/l)}\=d\sum_{\cal O} t_{\cal O}^{(l)}:{\cal
O}:,\eqno(5.8)$$ gives $T_{n(/l)}$ in the {\it natural} operator decomposition.
Note that this procedure fixes the numerical distributions uniquely, up to the
normalization in the causal splitting $d_{\cal O}^{(l)}=r_{\cal O}^{(l)}-a_{\cal
O}^{(l)}$.

Starting with the natural operator decomposition of $D_n$ ($T_n$ resp.) and
$D_{n/l}^\mu$ ($T_{n/l}^\mu$), we commute with $Q$ or take the divergence
$\d^{x_l}_ \mu$ according to (5.4) and obtain the {\it natural operator
decomposition} of (5.4) ((1.7) resp.).  However, due to (5.5), the Cg-identities
for $D_n$ cannot be proven directly by decomposing (5.4). We must go another
way: {\it Instead of proving the operator gauge invariance (1.7), we prove the
corresponding Cg-identities (by induction on $n$), which are a stronger
statement.} In this framework the Cg-identities for $D_n$ can be proven by means
of the Cg-identities for $T_k,\,\tilde T_k$ in lower orders $1\leq k \leq n-1$.

The Cg-identities for $T_n$ are obtained by collecting all terms in the natural
operator decomposition of (1.7) which belong to a particular combination $:{\cal
O}:$ of external field operators. By doing this the arguments of some field
operators must be changed by using $\delta$-distributions, i.e. by applying the
simple identity $$:B(x_i){\cal O}(X):\delta (x_i-x_k)...= :B(x_k){\cal O}
(X):\delta (x_i-x_k)... \eqno (5.9)$$ where $X\=d (x_1,x_2,...x_n)$ and ${\cal
O} (X)$ means the external field operators besides $B$.

We are now able to give a precise definition of the statement that the
Cg-identities hold: {\it We start with the natural operator decomposition of
(1.7).  Using several times the identity (5.9), we can obtain an operator
decomposition $$[Q,T_n(X)]-i\sum_{l=1}^n\d_l T_{n/l}(X)=\sum_j\tau_j(X):{\cal
O}_j(X):, \eqno(5.10)$$ (where $\tau_j(X)$ is a numerical distribution and
$:{\cal O}_j(X):$ a normally ordered combination of external field operators)
which fulfils $$\tau_j(X)=0,\>\>\>\>\>\>\forall j.\eqno(5.11)$$ The
decomposition (5.10) must be invariant with respect to permutations of the
vertices.}

A Cg-identity is uniquely characterized by its operator combination $:{\cal
O}:$.  The terms in a Cg-identity are singular of order [11] $$|{\cal O}|+1
\eqno(5.12)$$ at $x=0$, where $$|{\cal O}|= 4-b-g_u-g_{\tilde u}-d.\eqno(5.13)$$
Here, $b,g_u,g_{\tilde u}$ are the number of gluons and ghost operators
$u,\tilde u$, respectively, in ${\cal O}$, and $d$ is the number of derivatives
on these field operators.

There are no pure vacuum diagrams contributing to (1.7), i.e. terms with no
external legs.  The disconnected diagrams fulfil the Cg-identities separately.
This can be proven easily by means of the Cg-identities for their connected
subdiagrams, which hold by the induction hypothesis.

Let us consider a {\it connected} diagram in the natural operator decomposition
of (1.7).  We call it {\it degenerate}, if it has at least one vertex with two
external legs; otherwise it is called {\it non-degenerate}. Let $x_i$ be the
degenerate vertex with two external fields, say $B_1,\,B_2$. Such a 'degenerate
term' has the following form
$$:B_1(x_i)B_2(x_i)B_3(x_{j_1})...B_r(x_{j_{r-2}}):\Delta (x_i-x_k)t_{n-1}
(x_1-x_n,...\overline{x_i-x_n},...x_{n-1}-x_n),\eqno(5.14)$$ where $k\not=
i,\,j_l\not=i\,(\forall l=1,...,r-2)$ and the coordinate with bar in $t_{n-1}$
must be omitted. In general, there is a sum of such terms (5.14) belonging to
the fixed (degenerate) operator combination $:{\cal
O}:=:B_1(x_i)B_2(x_i)B_3(x_{j_1})...B_r(x_{j_{r-2}}):$.  For $\Delta (x_i-x_k)$
the following possibilities appear:

(a) $\Delta =D_F,\,\d D_F,\,\d_\mu\d_\nu D_F\,(\mu\not=\nu),\,
\d_\rho\d_\mu\d_\nu D_F\,(\mu\not=\nu\not=\rho\not=\mu),$

(b) $\Delta =\delta^{(4)},\,\d\delta^{(4)}$.

The $\d\delta^{(4)}$-terms in (b) cancel [15].  If a degenerate term (5.14) with
$\Delta =\delta^{(4)}$ (type (b)) can be transformed in a non-degenerate one by
applying (possibly several times) the identity (5.9) only, we call it {\it
$\delta$-degenerate}; if this is not possible we call it {\it truly degenerate}.
All other degenerate terms (i.e. the terms of type (a)) are called truly
degenerate, too.

{\it The truly degenerate terms fulfil the Cg-identities separately, by means of
the Cg-identities for their subdiagrams} (sect.3.1 of [19]). The latter hold by
the induction hypothesis. The exception are some tree diagrams in second and
third order, which need an explicit calculation (sect.3.2 of [19]).

There remain the non-degenerate and $\delta$-degenerate terms, which are
linearly dependent.  Therefore, {\it the $\delta$-degenerate terms must be
transformed in non-degenerate form by using (5.9)}. In this way we obtain
completely {\it new} Cg-identities, in contrast to the disconnected and the
truly degenerate Cg-identities, which rely on Cg-identities in lower orders.
Therefore, it is not astonishing that the difficult part of the proof of the
Cg-identities concerns the non-degenerate $:{\cal O}:$ (including
$\delta$-degenerate terms). First one proves the Cg-identities of the
non-degenerate and $\delta$-degenerate terms for $A_n^\prime, R_n^\prime$ (and
therefore also for $D_n$) by means of the Cg-identities in lower orders
(sect.4.1 of [19]). In the process of distribution splitting the Cg-identities
can be violated by local terms only which are singular of order $|{\cal O}|+1$
(5.12), i.e.  the possible anomaly has the form
$$a(x_1,...x_n)=\sum_{|b|=0}^{|{\cal O}|+1}C_b D^b\delta^{4(n-1)}(x_1-x_n,...).
\eqno(5.15)$$ We see that we only have to consider Cg-identities with $$|{\cal
O}|\geq -1.\eqno(5.16)$$ This occurs only for Cg-identities with 2-,3-,4-legs
and one Cg-identity with 5-legs\- ($:{\cal O}:=:uAAAA:$). For the latter the
colour and Lorentz structures exclude an anomaly (5.15) [15]. For the
Cg-identities with 2-,3- and 4-legs we first restrict the constants $C_b$ in the
ansatz (5.15) by means of covariance, the SU(N)-invariance and invariance with
respect to permutations of the inner vertices.  Then we remove the possible
anomaly by finite renormalizations of the $t$-distributions in the Cg-identity.
If a certain distribution $t$ appears in several Cg-identities, the different
normalizations of $t$ must be compatible. For certain Cg-identities ($:{\cal
O}:=:uAA:,:uAAA:,:uu\d\tilde u A:$) the removal of the anomaly is only possible,
if one uses additional information about the infrared behavior of the
divergences with respect to inner vertices [15].  \vskip 0.5cm {\it 3. The
modifications of the proof of gauge invariance for arbitrary $\lambda$} \vskip
0.5cm Going over to an arbitrary $\lambda$-gauge there are two fundamental
changes:

(A) The wave equation for the free gauge field $A^\mu_a$ is replaced by (2.1).
However in the proof of gauge invariance the equation of motion for $A_\mu$ is
used in (5.1) only. Therefore, by working always with $\{Q,\d_\mu\tilde
u_a\}=i\d^\nu F_{a\nu\mu}$ the modification of the equation of motion causes no
changes in the proof of gauge invariance.

(B) The commutator $[A_\mu,A_\nu]$ (2.15) has an additional $\lambda$-dependent
term with the dipole distribution $E$.  Similar changes appear in the positive
and negative frequency part of (2.15), as well as in the retarded, advanced and
Feynman propagator. All other commutators rsp. propagators are independent of
$\lambda$, e.g. $[A_{a\mu},F_{b\nu\tau}]$. If we would work with another ghost
coupling $(\beta_1,\beta_2)\not= (0,0)$ the field $\d^\mu A_{\mu}$ would appear,
which has s a $\lambda$-dependent commutator with $A_{\nu}$ $$[\d^\mu
A_{a\mu}(x),A_{b\nu}(y)]={i \over \lambda} \delta_{ab} \d_\nu D(x-y).
\eqno(5.17)$$ We now have to check that the explicit form of the $AA$-commutator
rsp.  propagator is not used in the proof of the Cg-identities:

- Second order tree diagrams: The explicit form of the propagators is used in
the verification of gauge invariance for the second order tree diagrams.  But
gauge invariance can only be violated by local terms $\sim(\d)\delta (x_1-x_2)$.
The latter can only appear if the propagator is of singular order $\omega\geq
-1$ (see (5.12)). But the $AA$-propagator (without derivatives) has $\omega =
-2$ and, therefore, plays no role in this calculation.  In all other propagators
(with derivatives) the $\lambda$-dependence drops out because the derivatives
occur in the antisymmetric $F$, only.  Especially we conclude that the
four-gluon interaction (which is a normalization term of the second order tree
diagram with external legs $:A(x_1)A(x_1)A(x_2)A(x_2):$ and is uniquely fixed by
gauge invariance [11,19]) is independent of $\lambda$ and that it is the only
local term in $T_2^\la\vert_{\rm tree}$.

- $\delta$-degenerate terms: If $\Delta (x_i-x_k)$ in (5.14) originates from an
$AA$-propagator (without derivatives), we know about the singular order $\omega
(\Delta) \leq\omega([A,A])+1=-1$. Therefore, $\Delta\not= \delta,\d\delta$ and
the set of $\delta$-degenerate terms is unchanged for $\lambda\not= 1$.  Of
course most $t$-distributions depend on $\lambda$ (due to (2.15)), but we
conclude that the {\it Cg-identities belonging to non-degenerate $:{\cal O}:$}
(which include the $\delta$-degenerate terms) {\it are manifestly independent of
$\lambda$}. (This is obvious for the non-degenerate terms.)

- We turn to the proof of the Cg-identities belonging to non-degenerate $:{\cal
O}:$ for $a'$ and $r'$ by means of the Cg-identities in lower orders (sect.4.1
of [19]). There one has to show that the operator decomposition of
$[Q,A'_n]=[Q,\sum\tilde T_k T_{n-k}]$ is unchanged if we interchange the
operation $[Q,.]$ with contracting. For this purpose one needs the explicit form
of some propagators, but the $AA$-propagator is not used. The non-trivial step
is the cancellation of the terms arising by contracting the commutated leg.

- The same cancellation is used in the proof of the Cg-identities for the truly
degenerate terms by means of the Cg-identities for their subdiagrams (sect.3.1
of [19]). Again the explicit form of the $AA$-propagator plays no role.

We emphasize that (A) and (B) are the only relevant changes for arbitrary
$\lambda$. Especially the singular order of the numerical distributions
(5.12-13) and the symmetries (Lorentz covariance, SU(N)-invariance, P-,T- and
C-invariance, pseudo-unitarity and invariance with respect to permuatations of
the vertices) are manifestly independent of $\lambda$.  Consequently, the ansatz
(5.15) for the possible anomalies (in the Cg-identities belonging to
non-degenerate $:{\cal O}:$) remains the same and the constants $C_b$ in (5.15)
can be restricted in the same way. Moreover, the normalization polynomials of
the $t$-distributions are unchanged and, therefore, we can use them to remove
the anomalies in the same way. Finally gauge invariance of third order tree
diagrams, which must be verified explicitly (sect.3.2 of [19]), and the proof of
the non-trivial 5-legs Cg-identity [15] rely on the SU(N)-invariance and Lorentz
covariance. Therefore, also this parts of the proof need no change.

Summing up we see that the inductive proof of the Cg-identities is manifestly
independent of $\lambda$ if we choose the ghost coupling $\beta_1=0=\beta_2$
(5.3) and always work with $F_{\mu\nu}$ instead of $\d_\mu A_\nu$ (5.1).

The coupling to fermionic matter fields (in the fundamental representation) can
be added to this model. Gauge invariance holds true if and only if the coupling
constants agree (universality of charge). This has been carried out in the
Feynman gauge in [20]. There are no changes for arbitrary values of $\lambda$.

\appendix\section*{Appendix 1: Cauchy problem for the iterated wave equation}

First we formulate the Cauchy problem for the equation $$\w^2u\equiv
(\d_0^2-\d_1^2-\d_2^2-\d_3^2)^2u=0.\eqno(A.1)$$ Since (A.1) is of forth order in
time $x_0=t$, a complete set of Cauchy data at $t=0$ is given by $$(\d_0^n
u)(0,\vec x)=u_n(\vec x),\quad n=0,1,2,3.\eqno(A.2)$$ For simplicity we assume
the $u_n$ to be in Schwartz space, then the initial-value problem (A.1) (A.2)
has a unique solution. This solution can be constructed by means of the tempered
distributions $D(x)$ and $E(x)$, defined by $$\w D=0,\quad D(0,\vec x)=0,\quad
(\d_0D)(0,\vec x)=\delta^3(\vec x), \eqno(A.3)$$ $$\w^2E=0,\quad
(\d_0^nE)(0,\vec x)=0,\quad n=0,1,2,\quad (\d_0^3E)(0,\vec x)=\delta^3(\vec
x).\eqno(A.4)$$ $D$ is the well known Pauli-Jordan distribution and $E$ is
sometimes called dipole distribution and we will soon compute it.

We now claim that the solution of the Cauchy problem (A.1) (A.2) is given by
$$u(x)=\int d^3y\,\Bigl[D(x-y)u_1(\vec y)-\d_0^yD(x-y)u_0(\vec y)+$$
$$+E(x-y)(u_3-\lap u_1)(\vec y)-\d_0^yE(x-y)(u_2-\lap u_0)(\vec y)
\Bigl],\eqno(A.5)$$ where $\lap$ denotes the three-dimensional Laplace operator.
This formula is the same as the covariant equation (2.6) which is an obvious
generalization of the solution of the ordinary wave equation. Using (A.3) and
(A.4) it is a simple task to verify (A.1) and (A.2).  Therefore it remains to
construct the dipole distribution $E$.

>From (A.3) and (A.4) we get $$\w E(x)=D(x)\eqno(A.6)$$ and we want to obtain $E$
as solution of this equation. We solve this problem in momentum space. The
Fourier transform of $D$ is well known $$\hat D(p)={i\over 2\pi}\sgn
p_0\delta(p^2),\eqno(A.7)$$ so that $$p^2\hat E(p)=-{i\over 2\pi}\sgn
p_0\delta(p^2).\eqno(A.8)$$ A solution of this equation can immediately be
written down by means of the identity $$p^2\delta'(p^2)={d\over
dp^2}\Bigl(p^2\delta(p^2)\Bigl)-\delta(p^2) =-\delta(p^2),\eqno(A.9)$$ namely
$$\hat E(p)={i\over 2\pi}\sgn p_0\delta'(p^2).\eqno(A.10)$$ By inverse Fourier
transform the initial conditions (A.4) can be verified and $E(x)$ can be
computed $$E(x)={1\over 8\pi}\sgn (x_0)\Theta(x^2).\eqno(A.11)$$ Note that the
positive frequency part $$"\hat E^{(+)}(p)"={i\over
2\pi}\Theta(p_0)\delta'(p^2)$$ is ill-defined. This never occurs in rigorous
calculations. Only derivatives of $E$ have to be split into positive and
negative frequency parts (see (1.3)) and these are well-defined.

\appendix\section*{Appendix 2}

Here we prove the relation (4.14). We start from the orthogonal direct
decomposition (4.9) which can be written as $${\bf 1}=P_Q+P_{Q^+}+P\eqno(A.12)$$
where $P_Q$ and $P_{Q^+}$ are projection operators onto $\overline{ {\rm Ran}
Q}$ and $\overline{{\rm Ran}Q^+}$ and $P$ projects on $\cal H_{\rm phys}$. The
operator (4.11) $$\{Q,Q^+\}\equiv K>0\eqno(A.13)$$ is positive selfadjoint on
the orthogonal complement $\cal H^\perp _{\rm phys}$ of $\cal H_{\rm phys}$, so
that it has an inverse $$KK^{-1}=P_Q+P_{Q^+}=K^{-1}K,\quad
K^{-1}P=0.\eqno(A.14)$$ This allows to write (A.12) in the form $${\bf
1}=P+QQ^+K^{-1}+Q^+QK^{-1}.\eqno(A.15)$$ Now we consider
$$PT(X_1)T(X_2)P=PT(X_1)(P+QQ^+K^{-1}+Q^+QK^{-1})T(X_2)P$$
$$=PT(X_1)PT(X_2)P+PT(X_1)QQ^+K^{-1}T(X_2)P+$$
$$+PT(X_1)Q^+QK^{-1}T(X_2)P.\eqno(A.16)$$ Since $PQ=0$, the second term is equal
to $$P[T(X_1),Q]Q^+K^{-1}T(X_2)P$$ which is a divergence due to gauge invariance
of $T(X_1)$.

In the last term in (A.16) we use the fact that $K$ and, hence, $K^{-1}$ commute
with $Q$ which follows easily from the definitions (A.13), (4.11) and (4.6).
Then we conclude that $$PT(X_1)Q^+K^{-1}QT(X_2)P=PT(X_1)Q^+K^{-1}[Q,T(X_2)]P$$
is also a divergence. Consequently, $$PT(X_1)T(X_2)P=PT(X_1)PT(X_2)P+{\rm div}$$
which is the desired relation (4.14).

\vskip 1cm {\it References}\vskip 1cm

[1] Becchi C, Rouet A, Stora R 1976 {\em Ann. Phys. \bf {98}} 287

[2] Joglekar S D, Lee B W 1976 {\em{Ann. Phys. \bf{97}}} 160

[3] Taylor J C 1971 {\em{Nucl. Phys. B \bf{33}}} 436

[4] Slavnov A A 1972 {\em{Theor. Math. Phys. \bf{10}}} 99

[5] Lee B W, Zinn-Justin J 1973 {\em{Phys. Rev. D \bf{7}}} 1049

[6] Kennedy D C, Lynn B W, Im C J - C Im, Stuart R G 1989

\hskip 0.5cm {\em{Nucl. Phys. B \bf{321}}} 83

[7] Kennedy D C, Lynn B W 1989 {\em{Nucl. Phys. B \bf{322}}} 1

[8] Kuroda M, Moultaka G, Schildknecht D 1991 {\em{Nucl. Phys. B \bf{350}}} 25

[9] Denner A, Dittmaier S, Weiglein G, hep-th/9505271

[10] Scharf G, Finite Quantum Electrodynamics: the causal approach,

\hskip 0.5cm second edition, 1995 Springer Verlag

[11] D\"utsch M, Hurth T, Krahe F, Scharf G 1993 {\it Nuovo Cimento {\bf 106 A}}
1029,

\hskip 0.5cm 1994 {\bf 107 A} 375

[12] Epstein H, Glaser V 1973 {\it Ann.Inst.Poincar\'e A} {\bf 29} 211

[13] D\"utsch M, Scharf G, hep-th/96 12 091

[14] Aste A, D\"utsch M, Scharf G 1997 {\it J. Phys. A} {\bf 30} 5785

[15] D\"utsch M, Hurth T, Scharf G 1995 {\it Nuovo Cimento {\bf 108 A}} 679,
1995 {\bf 108 A} 737

[16] Lautrup B 1967 {\it Mat. Fys. Medd. Dan. Vid. Selsk. {\bf 35}}, no. 11

[17] Bognar J, Indefinite Inner Product Spaces, 1974 Springer Verlag

[18] Galindo A 1962 {\it Comm. Pure Appl. Math.} {\bf 15} 423

[19] D\"utsch M 1996 {\it Nuovo Cimento {\bf 109 A}} 1145

[20] D\"utsch M 1996 {\it J. Phys. {\bf A 29}} 7597

[21] Hurth T, hep-th/95 11 139

\section*{Figure Caption}

Figure 1. Relation between Fock spaces with different values of the gauge
parameter $\lambda$.

\end{document}